\documentclass[conference]{IEEEtran}
\usepackage[T1]{fontenc}
\usepackage[latin9]{inputenc}
\usepackage{color}
\usepackage{float}
\usepackage{amsthm}
\usepackage[nosumlimits]{amsmath}
\usepackage{amssymb}
\usepackage{multirow}
\usepackage{overpic}

\makeatletter

\floatstyle{ruled}
\newfloat{algorithm}{tbp}{loa}
\providecommand{\algorithmname}{Algorithm}
\floatname{algorithm}{\protect\algorithmname}

\theoremstyle{plain}

\theoremstyle{definition}

\theoremstyle{plain}

\theoremstyle{remark}

\theoremstyle{remark}

\allowdisplaybreaks

\usepackage{mathtools}

\usepackage{amsfonts}
\usepackage{cite}
\usepackage{array}
\usepackage{algorithm}
\usepackage{algorithmic}
\usepackage{graphicx}
\newcommand*{\rom}[1]{\expandafter\@slowromancap\romannumeral #1@}

\makeatother

\providecommand{\definitionname}{Definition}
\providecommand{\factname}{Fact}
\providecommand{\remarkname}{Remark}
\providecommand{\theoremname}{Theorem}
\providecommand{\lemmaname}{Lemma}

\mathchardef\mhyphen="2D

\usepackage{siunitx} 
\usepackage{subfig}
\usepackage[font = footnotesize]{caption}

\usepackage{url}
\usepackage{hyperref}

\usepackage{booktabs}
\usepackage{multirow}
\usepackage{graphicx}

\begin{document}

\title{On the Potential of Using Sub-THz Frequencies for Beyond 5G}

\author{
\IEEEauthorblockN{Oskari Tervo\textsuperscript{$\ast$}, Ilmari Nousiainen\textsuperscript{$\dag$}, Ismael Peruga Nasarre\textsuperscript{$\ast$}, Esa Tiirola\textsuperscript{$\ast$}, Jari Hulkkonen\textsuperscript{$\ast$} } %

\IEEEauthorblockA{\textsuperscript{$\ast$} Nokia, Oulu, Finland} 
\IEEEauthorblockA{\textsuperscript{$\dag$} University of Oulu, Oulu, Finland}
}

\maketitle




\begin{abstract}

This paper studies the potential of using above 71GHz frequencies for 5G-Advanced or later in 6G. More specifically, the focus is to analyze what could be needed in terms of waveform and numerologies. The results suggest that higher baseline subcarrier spacings (SCSs) may be needed when moving above 71GHz, to fulfill the need for higher required bandwidths and phase noise robustness. The required SCS depends on carrier frequency and modulation order. It is also illustrated that single-carrier waveforms, especially Known Tail Discrete Fourier Transform Spread Orthogonal Frequency Division Multiplexing (KT-DFT-s-OFDM) waveform is a potential candidate to be used in 5G-Advanced or 6G for sub-THz frequencies due to its robustness to phase noise, lower output power back-off and flexible adaptation of head and tail lengths.

\end{abstract}


\section{Introduction}
Cellular technologies need to enter into new spectrum bands to cope with increased capacity requirements. Sub-THz will be the next step to expand from mm-waves towards higher bandwidths and thus enable multi-Gbps and up to Tbps data rates that future immersive multimedia experiences like eye-resolution extended reality (XR), holograms, etc. will require. Sub-THz spectrum bands is able to cope with very high area capacity in the traffic hot spots with short range communication and including device to device. Equally important is long range fixed and mobile wireless links for access and backhaul. Moreover, sub-THz frequencies can enable new types of use cases and properties such as integrating sensing to communications and extended reality \cite{HexaD2.1}. Especially joint communication and sensing could be considered as part of 6G from the beginning, and huge bandwidths offered by sub-THz region may provide high-resolution sensing opportunities \cite{Wild-21}. Spectrum regulators have made effort to enable expansion to THz bands. The federal communications commission (FCC) has decided to adopt new rules for the bands above 95 GHz and it is already  possible to get experimental licenses to use frequencies beyond 95 GHz \cite{FCC:SpectrumHorizons}.

5G New Radio (NR) standards release (Rel) 15 and Rel-16 support carrier frequencies up to \SI{52.6}{GHz}. Rel-17 has been already standardizing the first version for frequency range (FR) 52.6-71GHz (i.e., FR2-2), which is a quite straightforward extension from Frequency Range 2-1 (FR2-1) to enable fast time-to-market \cite{RP-202925}. It is expected that above 71GHz frequencies will be considered for the future 5G-Advanced releases and later in 6G.   

Going to above 71GHz frequencies, the transceiver impairments such as phase noise (PN) and power amplifier (PA) distortions will be further increasing, which result in coverage challenges due to large output power back-off required for the power amplifiers \cite{Tervo-20}. This makes the applied waveform of uttermost importance, and single-carrier like waveforms may be preferred to alleviate these challenges. Another important note is that to achieve the real potential of sub-THz frequencies to provide Tbps data rates, the bandwidths need to be huge. There are basically few options to provide huge bandwidths in OFDM-based systems as 5G New Radio: 1) use smaller subcarrier spacings (SCSs) with larger Fast Fourier Transform (FFT) size than the current 4096; 2) use smaller SCSs with 4096 FFT size, and use carrier aggregation; 3) Use large SCS and 4096 FFT size. Approach 1) is complex due to large Inverse FFT (IFFT) and FFT operations, and the phase noise is problematic due to severe inter-carrier interference. Approach 2) is not power efficient because use of carrier aggregation will destroy the single carrier properties of the signal, resulting in high peak to average power ratio (PAPR) and thus lower coverage. Furthermore, smaller SCSs are not robust in case of high phase noise. On the other hand, approach 3) would keep the existing FFT complexity, PAPR benefits, and smaller impact of phase noise. Thus, in this paper we focus on this approach adopted also in 52.6-71GHz, where the scalable numerology would remain the same otherwise, except the SCSs would be increased.

In 3GPP Rel-18 workshop, and in the following email discussion prior to RAN\#93-e  already, many companies were proposing single-carrier waveform study for >71 GHz. However, the discussions lead to the conclusion that the studies will be postponed, and may begin in Rel-19 or beyond. The frequency bands in this region include especially W-band (75 to 110  GHz) but also D-band (110 to 170 GHz) \cite{Xing-21}. As illustrated e.g., in  \cite{Xing-21},  the amount of spectrum is high, but there are also restrictions for the band usage, such as RR5.340 where all emissions are prohibited (passive satellite band).
European telecommunications regulator CEPT ECC has approved two recommendations for Fixed Service (FS) above 92 GHz \cite{ECC-01,ECC-02}:
\begin{itemize}	
\item W Band ECC Recommendation ECC/REC/(18)02 on frequencies 92-114.25 GHz
\item D Band ECC Recommendation ECC/REC/(18)01 on frequencies 130-174.8 GHz.
\end{itemize}
Thus, W and D bands may be primarily allocated for mobile services together with fixed services. Nevertheless, based on current situation, continuous bands for mobile/fixed services up to 275GHz range roughly from 2GHz to 20 GHz. There is currently an on-going work in ITU-R WP5D to create a report for the technical feasibility of IMT in bands above 100GHz. This is aiming to provide a recommendation for the usage of the bands, to address the characteristics and benefits of above 100GHz bands in WRC-23.


To demonstrate the potential of above 71GHz frequencies in 5G Advanced or 6G, this paper studies physical layer numerology and waveforms for beyond \SI{71}{GHz} and sub-THz communications. 
We will focus on the Known Tail (KT)-DFT-s-OFDM waveform, and compare it to the currently supported waveforms Cyclic Prefix (CP)-OFDM and DFT-s-OFDM in 5G NR Rel-15. First the significant difference in the achievable PA output power with different waveforms is demonstrated, showing the importance of supporting single carrier waveforms. Then, the link performance of is investigated, taking into account the phase noise,  currently supported SCSs, and also various higher SCS options, to find out the required numerology to enable high throughput communications in sub-THz frequencies.

\section{Design Principles for sub-THz Operation}
\label{sec:5GNowAndBeyond71GHz}


\begin{table*}[t!]
\centering
\vspace{4mm}
\caption{Numerology scaling framework according to 5G New Radio.}
\label{tab:Numerology}
\begin{tabular*}{\textwidth}{@{\extracolsep{\fill}}ccccccccccc@{\extracolsep{\fill}}}
\toprule
\multicolumn{1}{l}{} &
  \multicolumn{1}{l}{} &
  \multicolumn{1}{l}{} &
  \multicolumn{1}{l}{} &
  \multicolumn{2}{c}{\textbf{\begin{tabular}[c]{@{}c@{}}CP length\\ 14 symbols/slot\end{tabular}}} &
  \textbf{\begin{tabular}[c]{@{}c@{}}CP length ECP\\ 12 symbols/slot\end{tabular}} &
  \textbf{\begin{tabular}[c]{@{}c@{}}CP length\\ (no-CP)\\ 15 \\ symbols/slot\end{tabular}} &
  \textbf{} &
  \multicolumn{2}{c}{\textbf{BW 4096-size FFT}} \\ \cmidrule(l){5-11} 
\textbf{$\mu$} &
  \textbf{$\Delta f$} &
  \textbf{\begin{tabular}[c]{@{}c@{}}Symbol \\ length\\ {[}$\mu s${]}\end{tabular}} &
  \textbf{\begin{tabular}[c]{@{}c@{}}Sample \\ rate\\ {[}Ms/s{]}\end{tabular}} &
  \textbf{\begin{tabular}[c]{@{}c@{}}Regular CP\\ {[}$\mu s${]}\end{tabular}} &
  \textbf{\begin{tabular}[c]{@{}c@{}}Special CP\\ {[}$\mu s${]}\end{tabular}} &
  \textbf{\begin{tabular}[c]{@{}c@{}}Extended CP\\ {[}$\mu s${]}\end{tabular}} &
  \textbf{{[}$\mu s${]}} &
  \textbf{\begin{tabular}[c]{@{}c@{}}Slot \\ duration\\ {[}$\mu s${]}\end{tabular}} &
  \textbf{\begin{tabular}[c]{@{}c@{}}Channel \\ BW \\ {[}GHz{]}\end{tabular}} &
  \textbf{\begin{tabular}[c]{@{}c@{}}OCB \\ (264 PRBs)\\ {[}GHz{]}\end{tabular}} \\ \midrule
0 & 15   & 66.67 & 61.44    & 4.6875 & 5.2083 & 16.667 & 0 & 1000     & 0.05 & 0.05  \\
1 & 30   & 33.33 & 122.88   & 2.3438 & 2.8646 & 8.333  & 0 & 500      & 0.1  & 0.10  \\
2 & 60   & 16.67 & 245.76   & 1.1719 & 1.6927 & 4.167  & 0 & 250      & 0.2  & 0.19  \\
3 & 120  & 8.33  & 491.52   & 0.5859 & 1.1068 & 2.083  & 0 & 125      & 0.4  & 0.38  \\
4 & 240  & 4.17  & 983.04   & 0.2930 & 0.8138 & 1.042  & 0 & 62.5     & 0.8  & 0.76  \\
5 & 480  & 2.08  & 1966.08  & 0.1465 & 0.6673 & 0.521  & 0 & 31.25    & 1.6  & 1.52  \\
6 & 960  & 1.04  & 3932.16  & 0.0732 & 0.5941 & 0.260  & 0 & 15.625   & 3.2  & 3.04  \\
7 & 1920 & 0.52  & 7864.32  & 0.0366 & 0.5575 & 0.130  & 0 & 7.8125   & 6.4  & 6.08  \\
8 & 3840 & 0.260 & 15728.64 & 0.0183 & 0.5391 & 0.065  & 0 & 3.90625  & 12.8 & 12.17 \\
9 & 7680 & 0.130 & 31457.28 & 0.0092 & 0.5300 & 0.033  & 0 & 1.953125 & 25.6 & 24.33 \\ \bottomrule
\end{tabular*}%
\end{table*}

\subsection{Physical Layer Numerology}

The 5G NR is designed to support wide range of SCSs to handle different use cases and a wide range of supported carrier frequencies. For the Frequency Range 2-2 (FR2-2), which is the frequency range 52.6-71GHz, Rel-17 has specified 120kHz SCS as the basic numerology (only mandatory), resulting in the baseline channel bandwidth of 400MHz. In addition to this, also 480kHz and 960kHz SCSs are supported, providing opportunity for up to 2GHz maximum channel bandwidth \cite{3GPPTS38211,R4-2202364}. Higher SCSs result in shorter cyclic prefix (CP), but on the other hand the delay spreads are decreasing in the higher frequencies due to propagation characteristics and beam-based operations \cite{R1-2008615}, so the CP overhead defined for lower bands with smaller SCS was still seen usable for these SCSs.

Following the same scalable numerology framework as illustrated in Table \ref{tab:Numerology}, it is expected that the 'baseline' SCSs should further be increased in above 71GHz in order to fulfill the need for phase noise robustness and huge bandwidths. When assuming the current baseline FFT size of 4096 and allocation bandwidths, SCSs from 960kHz to 3840kHz could provide bandwidths ranging from 3GHz to 12GHz for single carrier, which would already be quite well inline with the available continuous spectrum based on current regulation. Using this framework, CP length will decrease accordingly, but delay spread impact is expected to be decreased as well due to highly directional transmission. Also, to address this point a promising option could be to use waveform without CP extension, and use internal quard sequence before DFT to address the delay spread in a flexible manner. This is further discussed in the next sub-section.

\subsection{Waveforms}
Current 5G in FR2-2 is using the same waveforms as the lower frequency ranges (FR1 and FR2-1), i.e., CP-OFDM is used in downlink, while the uplink can use both CP-OFDM and DFT-s-OFDM. Uplink supports DFT-s-OFDM because it can provide significantly better coverage than CP-OFDM for coverage-limited user equipments (UEs). However, it is obvious that the coverage becomes even more challenging when going above 71GHz, and has to be one of the main design considerations there.

Regarding the waveform design, it is important to evaluate the potential needs in higher frequencies. First, it is obvious that the coverage issue drives towards a single-carrier waveform due to its lower PAPR characteristics compared to multi-carrier waveform \cite{leHang-22}. Also, it is expected that there are wide bandwidths available, giving opportunity to multiplex the users in time-domain or separate frequency bands, so that frequency multiplexing is not needed. Further, the expectation is that the systems will be mostly focused on analog/hybrid beamforming with low-order multiple-input multiple-output (MIMO) possibilities, giving already a restriction for user multiplexing. These aspects imply that single-carrier like waveforms are the most prominent options for sub-THz frequencies, also in downlink direction. 

\begin{figure}
     \centering
     \includegraphics[angle=0,width=0.95\columnwidth]{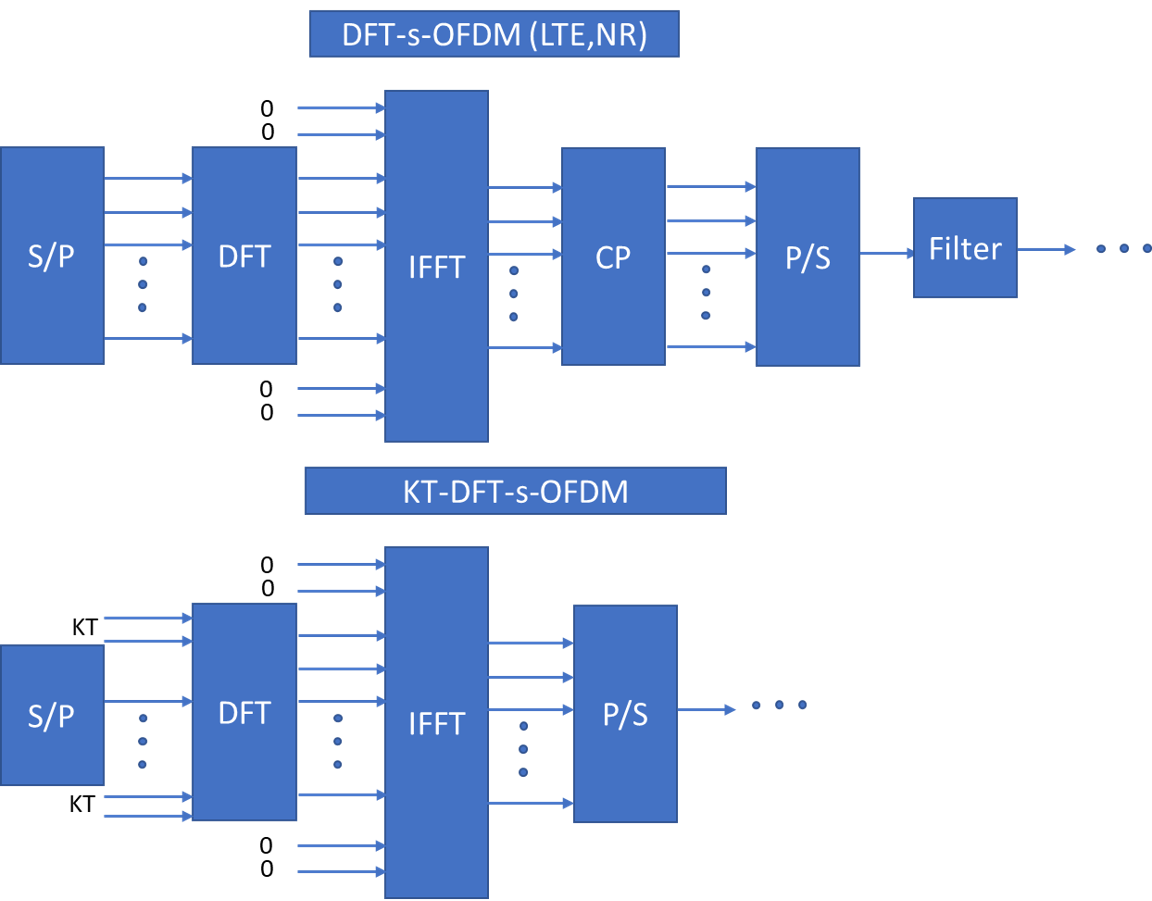}
     \caption{Illustration of DFT-s-OFDM and KT-DFT-s-OFDM waveforms.}
     \label{fig:WaveformBlocks}
 \end{figure}

\begin{figure}
     \centering
     \includegraphics[angle=0,width=0.95\columnwidth]{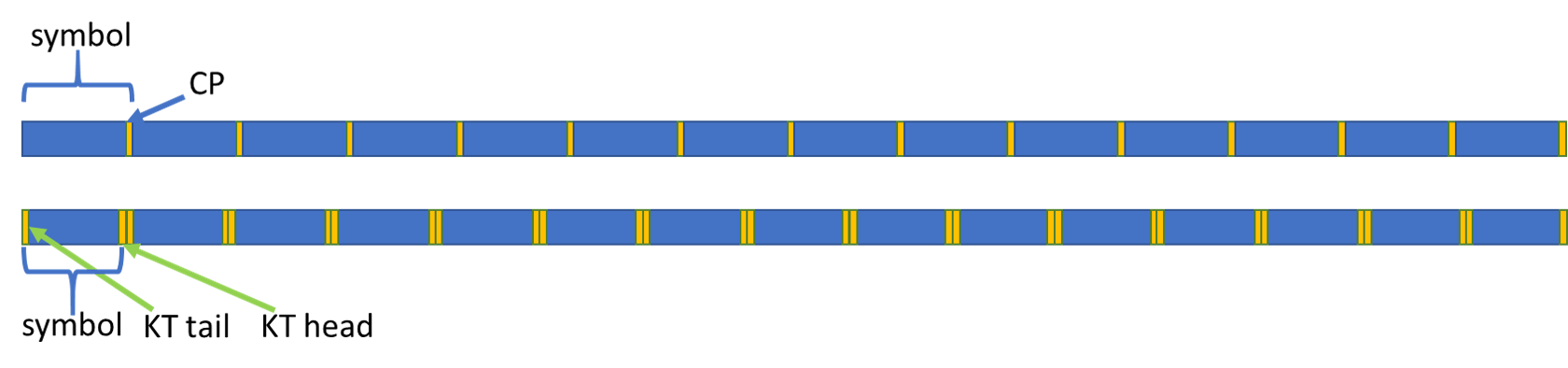}
     \caption{Comparison of CP and KT-based slots.}
     \label{fig:CPandKTslots}
 \end{figure}

Currently specified DFT-s-OFDM is fully based on OFDM with CP, but it employs DFT operation in the transmitter to spread the signal across the allocated subcarriers, resulting in lower PAPR. However, the restriction of this waveform is still CP which has relatively high overhead, especially if extended CP is used. Thus, it cannot provide flexibility to adapt different types of delay spreads and maximize the spectrum efficiency, because changing the CP length will change the slot structure. For example, 5G supports both normal and extended CPs. In case of normal CP, slot includes 14 symbols, while extended CP (ECP) results in 12 symbol slots, as also illustrated in Table \ref{tab:Numerology}. To this end, a promising candidate waveform is Known Tail (KT) DFT-s-OFDM (also known as Unique Word DFT-s-OFDM), which operates without actual CP extension \cite{Berardinelli-18}. It is fully based on legacy DFT-s-OFDM, but replaces the CP with in-symbol head (and tail) sequences appended prior to the DFT operation at the transmitter, as illustrated in Figure \ref{fig:WaveformBlocks}. The known sequences act in a similar manner as the CP, but the lengths of the sequences can flexibly adapt to different scenarios without affecting the slot structure, as illustrated in Figure \ref{fig:CPandKTslots}. This also results in lower emissions \cite{Berardinelli-18}. For KT sequence, e.g., Zadoff-Chu or $\pi/2$-BPSK can be used \cite{Berardinelli-18,Sahin-15}. To enable further coverage gains especially for low-order modulations, KT-DFT-s-OFDM can also insert frequency domain spectrum shaping with spectral extension operations after the DFT operations \cite{Peruga-21}.

\subsection{Power Amplifier Efficiency and Output Back-off}

Higher frequencies have many limitations which require careful study. The first major drawback requiring consideration is the decreased PA efficiency at higher carrier frequencies. For example, in \cite[Section 6.1.9.1]{3GPPTR38803}, it is shown that the output power of PAs for a given integrated circuit technology roughly degrades by \SI{20}{dB} per decade. This imposes a significant need to support waveforms and modulations that allow to achieve very low PAPR in order to achieve better power efficiency in base station (BS) and user equipment side, and to achieve the targeted maximum transmitted power levels. Maximizing the transmitter power is important because it directly translates into maximizing the cell coverage.

 \begin{figure}
     \centering
     \includegraphics[angle=0,width=0.99\columnwidth]{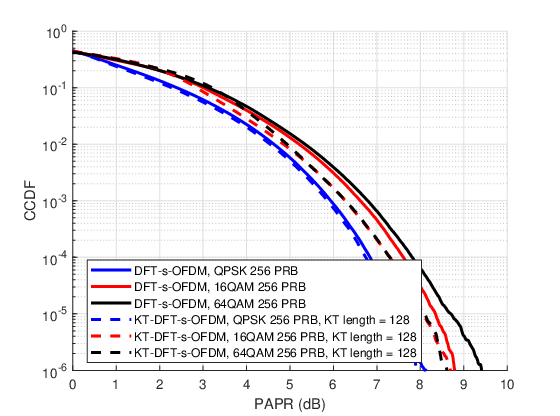}
     \caption{Comparison of PAPR between DFT-s-OFDM and KT-DFT-s-OFDM for KT length of 128, and [1+D]$\pi/2$-BPSK modulation for the KT sequences.}
     \label{fig:PAPR}
 \end{figure}

To allow for lower PAPR while maintaining the spectral efficiency, KT-DFT-s-OFDM can use KT sequences with near to constant envelope. The effects in the PAPR reduction with the same modulation used for data can be seen in Fig. \ref{fig:PAPR}, where a comparison between DFT-s-OFDM and KT-DFT-s-OFDM in terms of PAPR is performed, for KT length of 128 symbols and using [1+D]$\pi/2$-BPSK modulation (known for its low PAPR properties) for the KT sequences.

\begin{figure}
     \centering
     \includegraphics[angle=0,width=0.99\columnwidth]{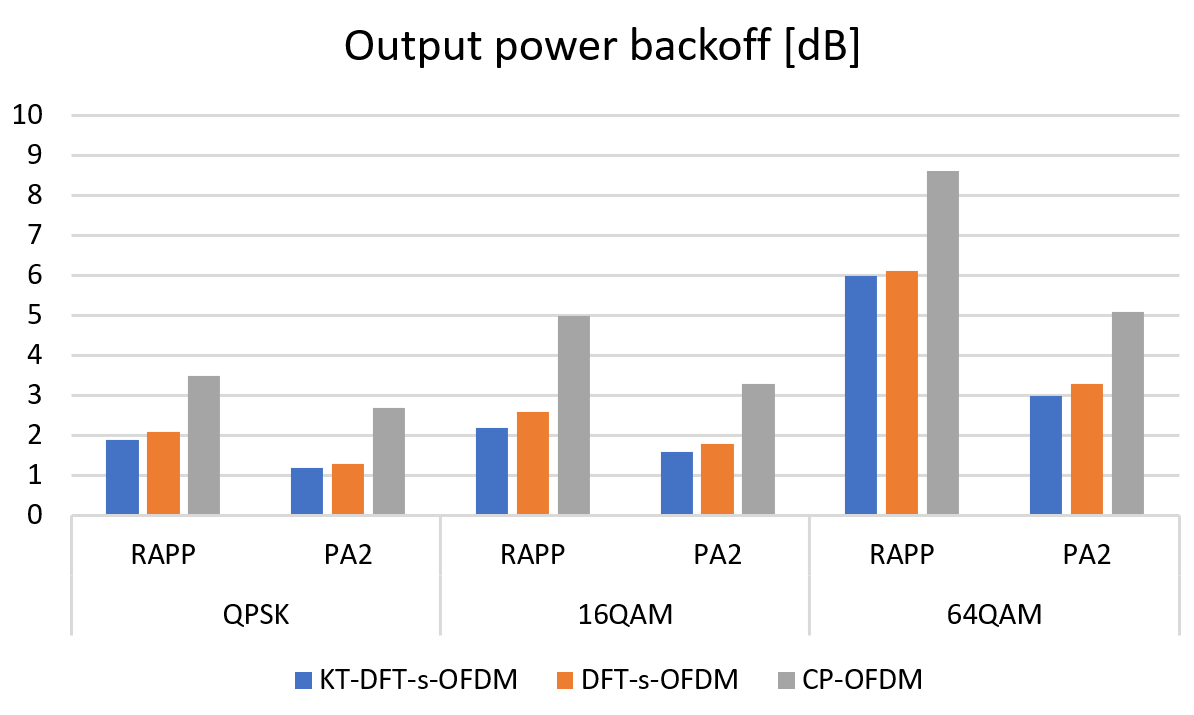}
     \caption{Comparison of PA output power back-off.}
     \label{fig:OBO}
 \end{figure}

Next, we evaluate the achievable output power. In Fig. \ref{fig:OBO}, the required output back-off (OBO) is shown for different waveforms, when assuming FR2-1 radio frequency (RF) requirements. In our simulations the RF requirements defined for FR2-1 in \cite{3GPPTS38101-2} are used since they correspond to the highest frequency bands which have been defined in 3GPP. More concretely, to obtain the OBO with respect to the 1 dB compression point of the PA for the different waveform/modulation, the signal is transmitted through a PA model, and the output is measured with respect to the RF emission limits defined for FR2-1 in \cite{3GPPTS38101-2}, the OBO is changed until all the requirements are met. Precisely, in-band emissions (IBE), error vector magnitude (EVM), adjacent channel leakage ratio (ACLR) and occupied bandwidth (OCB) are the considered RF emission requirements for the evaluation. Two PA models have been tested for the evaluation, corresponding to Rapp PA model from \cite{80211adRapp} and a second PA, denoted as PA 2, corresponding to a measured PA with working frequency beyond 140 GHz. It could be expected that some of the RF requirements (e.g., IBE, ACLR and OCB) would be relaxed when going to higher frequency bands, due to higher path loss and beamforming-based operation. The trend of relaxing RF requirements is already visible when comparing FR1 requirements \cite{3GPPTS38101-1} to FR2-1 \cite{3GPPTS38101-2}.

In Fig. \ref{fig:OBO}, it is observed that KT-DFT-s-OFDM can provide enhanced transmission power for each modulation when comparing to DFT-s-OFDM because of the lower PAPR and better behavior in frequency domain. Note that the OBO values shown in the figure are optimistic and obviously depends on the PA model, for both Rapp and PA 2 models, and could be clearly higher in practice, especially assuming that the required OBO will increase as a function of the bandwidth and carrier frequency. However, similar significant differences between the waveforms will anyway exist. The significant differences between CP-OFDM and DFT-s-OFDM can be already seen e.g., in allowed maximum power reduction values specified in \cite{3GPPTS38101-2} for different power classes in FR2.

\section{Radio Link Performance at sub-THz frequencies}
\label{sec:results}

In this section, the performance of KT-DFT-s-OFDM and DFT-s-OFDM waveforms are evaluated over different SCSs and carrier frequencies. We assume allocation of 256 physical resource blocks for each SCS, which follows using the maximum FFT size of 4096. By using SCSs from 120kHz to 3840kHz (i.e., varying $\mu$ between 4 and 8), the bandwidth thus varies from 400MHz to 12GHz.
We assume a tapped delay line A (TDL-A) with 5ns root-mean-squared (RMS) delay spread channel model, which is used also in 3GPP Rel-17 work item "Extending FR2 up to 71GHz". In all cases, a UE mobility of \SI{3}{km/h} is assumed. 
We use the PN model from \cite{R4-2010176} which scales according to the carrier frequency, and we use design margin 3dB for the PN model. The results assume perfect synchronization.

For reference signal configurations, we use 5G New Radio demodulation reference signal (DMRS) configuration type 1 and mapping type A, i.e., the DMRS is in the third symbol of a slot. For phase tracking reference signal (PTRS), we use maximum density i.e. 8x4 pattern for DFT-s-OFDM \cite{3GPPTS38211,Levanen-21}, to achieve the best possible performance. For KT-DFT-s-OFDM, we take only 6x4 PTRS groups, since KT sequences occupy the positions of the first and last PTRS groups. For KT-DFT-s-OFDM, where the CP is not used, we use 15 symbols per slot as was illustrated in Figure \ref{fig:CPandKTslots}, while legacy waveforms use normal 14 symbol slots. This is done to maintain the subframe duration of 1ms.



\begin{figure}[!t]
    \centering
        \subfloat{\includegraphics[width=1\columnwidth]{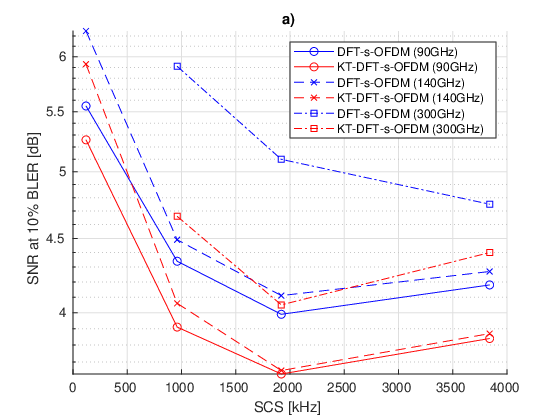}
        \label{fig:QPSK}}
        \hfil
        \subfloat{\includegraphics[width=1\columnwidth]{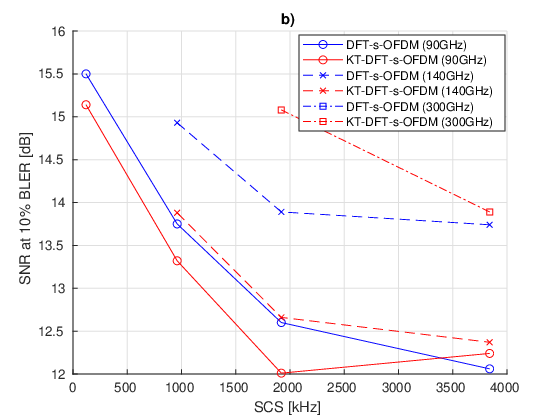}
        \label{fig:16QAM}}
        \hfil
        \subfloat{\includegraphics[width=1\columnwidth]{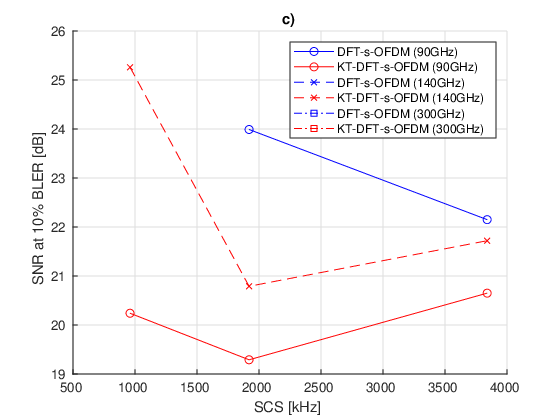}
        \label{fig:64QAM}}
    \caption{Link performance including OBO for different modulations, for (a) QPSK, (b) 16QAM and (c) 64QAM.}
    \label{fig:BLERandOBOresults}
\end{figure}


The coded link performance results with LDPC coding scheme following 5G NR specifications \cite{3GPPTS38211} are illustrated in Figs. \ref{fig:BLERandOBOresults}(a), \ref{fig:BLERandOBOresults}(b), and \ref{fig:BLERandOBOresults}(c) for QPSK (code rate 1/2), 16QAM (code rate 2/3) and 64QAM (code rate 2/3), respectively. Here we have chosen fixed KT length of 128, which gives approximately the same throughput in DFT-s-OFDM and KT-DFT-s-OFDM. The curves show signal to noise ratio (SNR) to achieve 10\% block error rate (BLER) including the required OBO according to PA 2 model, for carrier frequencies 90GHz, 140GHz and 300GHz. The required OBO for each combination is considered by scaling the SNR according to the results in Fig. \ref{fig:OBO} to fulfill the RF requirements. 

First it is observed that for each case, KT-DFT-s-OFDM is shown to outperform DFT-s-OFDM. Another important observation is that there is already clear performance difference between 120kHz and 960kHz SCSs even for QPSK, which implies that increase in the baseline SCS may be needed to reduce the PN impact. For QPSK, 960kHz SCS gives already good result up to 300GHz for KT-DFT-s-OFDM. However, DFT-s-OFDM with 300GHz suffers already 1.5dB performance loss for 960kHz compared to KT-DFT-s-OFDM. This reason of the loss stems mostly from better PN compensation capability of KT-DFT-s-OFDM, because the known sequences in head/tail can be used to track the PN in addition to the PTRS symbols.

When looking at 16QAM, it is observed that even 1920kHz SCS may be required, and KT-DFT-s-OFDM gives about 0.6dB performance gain over DFT-s-OFDM already at 90GHz carrier frequency. The gap increases when going to higher frequencies (1.5dB for 140GHz), and DFT-s-OFDM cannot achieve 10\% BLER target at 300GHz, while KT-DFT-s-OFDM can still do so.

On the other hand, for 64QAM DFT-s-OFDM can barely achieve 10\% BLER target for 90GHz carrier frequency if 1920kHz or 3840kHz SCS is used. KT-DFT-s-OFDM still works well for 960-3840kHz SCSs if 90GHz carrier frequency is used, and provides still reasonable results even for 140GHz, when 1920kHz or 3840kHz SCS is used. However, when 300GHz carrier frequency is used, the PN already becomes too significant in these cases even for KT-DFT-s-OFDM.

Based on the results it can be concluded that the higher SCSs from 960 to at least 3840kHz may be required when using carrier frequencies above 71GHz, and the required SCS is increasing according to the carrier frequency and modulation order. Another observation is that KT-DFT-s-OFDM is promising candidate waveform, providing robustness to phase noise, decreased OBO, and flexibly adjustable overhead for different scenarios.

\begin{figure}
     \centering
     \includegraphics[angle=0,width=1\columnwidth]{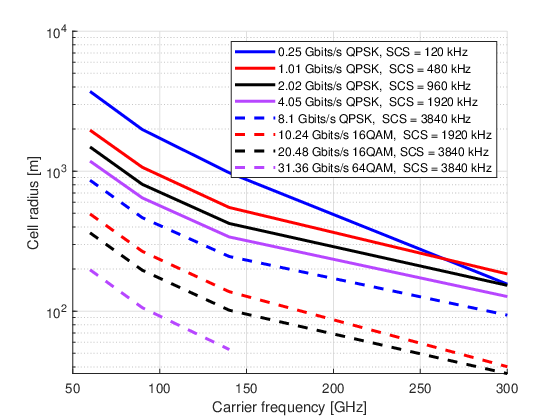}
     \caption{Example link budget assuming 60dBm EIRP and 256 active PRB for all cases.}
     \label{fig:LinkBudget}
 \end{figure}

Finally, Figure \ref{fig:LinkBudget} demonstrates the simple link budget assuming UMi LoS Street Canyon path loss model \cite{3GPPTR38901}, without atmospheric attenuation. Base station is assumed to have 16x16 array, while UE has 4x2 array. Effective Isotropic Radiated Power (EIRP) is assumed to be 60dBm. In the figure, different modulation orders are shown with the same code rates as in the earlier figures, and the higher the rate, the higher is the SCS and bandwidth. It is observed that for 140GHz carrier frequency, about 20Gbits/s rates can be achieved up to 100m e.g., assuming 3840kHz SCS with 16QAM modulation, and 2Gbits/s for up to 400m assuming 960kHz SCS with QPSK. On the other hand, coverage goes to about 3-5 times smaller at 300GHz depending on the rates.

\section{Conclusions}
\label{sec:conclusions}
This paper has studied the potential of using above 71GHz carrier frequencies in 5G Advanced or 6G. It has been demonstrated that KT-DFT-s-OFDM together with increased SCSs can be seen as promising direction to fulfil the needs for required large bandwidths and coverage in sub-THz frequencies. SCSs from 960kHz to 3840kHz can provide good baselines depending on the carrier frequencies and modulation order.

\bibliographystyle{IEEEtran}


\end{document}